\documentclass{ieeetj}
\usepackage{cite}
\usepackage{amsmath,amssymb,amsfonts}
\usepackage{graphicx,color}
\usepackage{textcomp}
\usepackage{xcolor}
\usepackage{hyperref}
\hypersetup{hidelinks=true}
\usepackage{algorithm,algorithmic}

\usepackage{tabularx}
\usepackage{pifont} 

\def\BibTeX{{\rm B\kern-.05em{\sc i\kern-.025em b}\kern-.08em
    T\kern-.1667em\lower.7ex\hbox{E}\kern-.125emX}}
\AtBeginDocument{\definecolor{tmlcncolor}{cmyk}{0.93,0.59,0.15,0.02}\definecolor{NavyBlue}{RGB}{0,86,125}}

\def\OJlogo{\vspace{-4pt}$<$Society logo(s) and publication title will appear here.$>$}
\def\seclogo{\vspace{10pt}$<$Society logo(s) and publication title will appear here.$>$}

\def\authorrefmark#1{\ensuremath{^{\textbf{#1}}}}

\begin{document}
\receiveddate{XX Month, XXXX}
\reviseddate{XX Month, XXXX}
\accepteddate{XX Month, XXXX}
\publisheddate{XX Month, XXXX}
\currentdate{XX Month, XXXX}
\doiinfo{XXXX.2022.1234567}


\title{Blind Evaluation Framework for Fully Homomorphic Encryption and Privacy-Preserving Machine Learning}

\author{Hunjae "Timothy" Lee\authorrefmark{1}, Corey Clark\authorrefmark{1}}
\affil{Southern Methodist University, Dallas, TX 75205 USA}
\corresp{Corresponding author: Hunjae Lee (email: hunjael@smu.edu).}
\authornote{This work was funded by BALANCED Media$\vert$Technology (BMT), a company that may potentially benefit from the research results. Dr. Corey Clark has an equity interest in BMT and also serves as the company's chief technology officer. The terms of this arrangement have been reviewed and approved by the Southern Methodist University in accordance with its conflict of interest policies.}

\begin{abstract}
In Privacy-Preserving Machine Learning (PPML), Fully Homomorphic Encryption (FHE) is often employed to enable computation directly on encrypted data. However, extensive programming with FHE have remained a challenge in large part owing to the difficulties in execution of control structures in encrypted state. While encrypted inference models are often immune from this issue due to their relative logical simplicity, training typically requires Interactive Rounds of Decryption and Evaluation (IRDE), also known as client-assisted computation, wherein certain operations are decrypted and evaluated in plaintext. However, such client-server communication with outsourced computing services cannot be reliably achieved in many scenarios such as in volunteer grids and distributed computing. These services have largely remained incompatible with PPML and FHE. To address this issue, we introduce the Blind Evaluation Framework (BEF), a cryptographically secure programming framework that enables encrypted execution of control structures and eliminates IRDE in training of PPML. To the best of our knowledge, we are the first to enable non-interactive training of PPML with FHE.
\\
As an example application of BEF, we implement an encrypted decision tree training model that doesn't require IRDE, a drastic improvement from the previous state-of-the-art which required $d$-rounds of IRDE for tree-depth $d$. By advancing the state-of-the-art in IRDE efficiency by eliminating IRDE entirely, BEF enables adoption of FHE in use-cases where ample computing resources are available without the ability for trusted clients to perform decryption rounds.
\end{abstract}



\maketitle

\section{INTRODUCTION}
\IEEEPARstart{F}{ully} Homomorphic Encryption (FHE) is an encryption scheme that allows computation on encrypted data without needing to decrypt them first, achieving encrypted computation without compromising cryptographic integrity \cite{gentry2009fully}. 
Due to this property, FHE is frequently used in the research of secure outsourced computing and Privacy-Preserving Machine Learning (PPML). However, there are limitations of FHE that create challenges for PPML. In \cite{gilad2016cryptonets,podschwadt2021non,lee2023optimizing}, encrypted predictive models are deployed for privacy-preserving neural networks but they are trained entirely in plaintext with FHE only being used for inference. In \cite{hesamifard2018privacy,brand2023practical}, encrypted training of PPML is achieved but requires that the untrusted computing party (server) engage in Interactive Rounds of Decryption and Evaluation (IRDE) with the private-key owner (client) to handle some operations in decrypted, plaintext form to aid the server. This is also called the client-assisted computation model, referring to the fact that while PPML with FHE is about secure outsourced computation, the client is often needed in regular intervals to handle operations that are challenging to handle in encrypted form. In this work, IRDE and client-assisted computation are used interchangeably. A similar method is employed in \cite{akavia2022privacy}, where a privacy-preserving decision tree training requires $d$-rounds of IRDE for tree-depth of $d$. These prior works that either require IRDE for encrypted training or ignore training completely are not the exceptions but rather the norm, with a more comprehensive list of prior works shown in Table \ref{tab:tableSummaryPriorWorks}. In this paper, we investigate the state of PPML with FHE and present a framework that enables non-interactive programming of training and inference models using FHE. Our contributions are as follows: we (1) present Blind Evaluation Framework (BEF) which enables truly encrypted programming without IRDE, and (2) demonstrate how BEF can eliminate IRDE in PPML training with a decision tree implementation.

\subsection{Fully Homomorphic Encryption (FHE)}
Mathematical proofs and definitions for FHE and relevant cryptographic functions from \cite{gentry2009fully} are summarized below.
\\
A homomorphic encryption scheme $\mathcal{E}$ is equipped with algorithms $KeyGen_{\mathcal{E}}$, $Encrypt_{\mathcal{E}}$, $Decrypt_{\mathcal{E}}$, and $Evaluate_{\mathcal{E}}$. In addition, the computational complexity of all of these algorithms must be polynomial in security parameter $\lambda$.
\\
The inputs to this scheme are: public key $pk$, a circuit $C$ from a permitted set $C_{\mathcal{E}}$ of circuits, and a tuple of ciphertexts $\Psi = <\psi_1,...,\psi_t>$.
\\
$\mathcal{E}$ is correct for circuits in $C_{\mathcal{E}}$ if, for any key-pair ($sk$, $pk$) output by $KeyGen_{\mathcal{E}}(\lambda)$, any circuit $C \in C_{\mathcal{E}}$, any plaintexts $\pi_1,...,\pi_t$, and any ciphertexts $\Psi = <\psi_1,...,\psi_t>$ with $\psi_i \gets Encrypt_{\mathcal{E}}(pk, \pi_i$), it is the case that:
\\\\
$\psi \gets Evaluate_{\mathcal{E}}(pk, C, \Psi) \Rightarrow C(\pi_1,...,\pi_t) = Decrypt_{\mathcal{E}}(sk,\psi)$
\\\\
Formal definition for homomorphic encryption is as following: \textit{$\mathcal{E}$ is homomorphic for circuits in $C_{\mathcal{E}}$ if $\mathcal{E}$ is correct for $C_{\mathcal{E}}$ and $Decrypt_{\mathcal{E}}$ can be expressed as a circuit $D_{\mathcal{E}}$ of size $poly(\lambda)$}. Furthermore, \textit{$\mathcal{E}$ is fully homomorphic if it is homomorphic for all circuits}. 

\subsection{Reality of PPML with FHE}
The current state of PPML with FHE is that non-interactive training has not been demonstrated. While non-interactive inference has been achieved a number of times on a variety of models (see section \ref{sec:related} for more details) due to their relative logical simplicity, few papers even attempt to perform encrypted training. Prior works that do perform encrypted training use IRDE to allow some operations to be handled in decrypted, plaintext form in a communication protocol between the server and client. While client-assisted protocols can be an acceptable compromise in some situations, they are inherently at odds with the spirit of FHE, given that the ultimate aim of FHE is to require decryption only once for final results. In addition, these architectures are not compatible in scenarios where a real-time connection with the client cannot be achieved reliably or if the client simply lacks bandwidth or computing power to stay online for the duration of computation. In short, the reality of PPML with FHE is that it has not yet delivered on the promises of \textit{fully} encrypted machine learning since portions of PPML are handled in plaintext using IRDE.

\subsection{Identifying the Problems}
Many prior works in PPML utilize integer or floating-point FHE libraries like CKKS \cite{cheon2017homomorphic} and BGV \cite{brakerski2014leveled} for their implementations. However, while these libraries allow arithmetic computation of encrypted data, they struggle with non-arithmetic programming such as handling conditionals, branching, decision making and more. For example, executing a simple if-statement returns an error because an if-statement inherently requires the evaluation and knowledge of the condition variable which is not possible when such condition is encrypted. Another example of a trivial process made challenging with FHE is comparison operations. While encrypted comparisons with FHE have been shown in prior works \cite{cong2022sortinghat,sun2018private,mahdavi2023level,frery2023privacy}, they were mostly used in limited contexts and have remained a challenge to implement in extensive encrypted programming because of their lack of usability as evaluation of the comparison result would require decryption. This is a notable deviation from conventional computing, where performing such if-statements or comparisons are routine and integral to various programs and algorithms. This issue persists beyond these examples and is prevalent in any programming process where evaluation of information is needed to perform further operations, such as with general control flows and decision statements.
\\
Because of these challenges, FHE is typically leveraged only for arithmetic computation while non-arithmetic, programmatic execution of logic and control structures are handled with IRDE. To put it simply, doing \textit{arithmetic computation} with FHE is trivial, but \textit{programming} with FHE is challenging.
\\
In privacy-preserving decision trees with FHE, \cite{cong2022sortinghat,mahdavi2023level,liang2024bpdte} used non-interactive comparison circuits with FHE but have only done so in limited contexts in decision tree inference without encrypted training. In training protocols where more complex programming logic is required, IRDE is often employed to resolve necessary operations in plaintext, on client-side. In \cite{akavia2022privacy}, feature selection during encrypted decision tree training was handled using IRDE, where the client would decrypt and select the correct feature to send it back to the server, resulting in $d$-rounds of IRDE for tree-depth of $d$. 
\\
In addition to IRDE, other approaches that address such limitations of FHE include using lookup tables \cite{boura2020chimera,lookuptable,cheon2024tree} and multi-party computation (MPC) \cite{de2014practical,du2002building,twin-cloud}. IRDE protocols with client-assisted computation, lookup table approaches, and MPC all rely on communication or computational involvement from the trusted client. IRDE protocols are differentiated from MPC approaches as IRDE models typically aim to reduce client involvement while MPC approaches rely on active and frequent interaction between two or more computing parties typically for not only computation efficiency but also for data privacy \cite{mpc_damgaard2012,mpc_cramer2001,du2001secure}.
\\
The above mentioned compromises and solutions can address different aspects of the limitations of FHE but can, in turn, create new limitations for the model or architecture of the system. The client-assisted computation model can be made efficient for the server, but require computational commitment from the client. This may not be feasible in cases where the client has little computational resources or cannot be guaranteed to stay online for the duration of the communication protocol. There are also scenarios where trusted clients cannot be made available for communication in real-time depending on the architecture of outsourced computation service. Lookup tables use pre-computed tables to replace functions difficult to evaluate in FHE and face similar constraints as with the client-server model as the lookup table approach requires computational commitment from the client.

\subsection{Alternative Representation of Programming Logic}
To address the challenges of programming with FHE, we use a Boolean FHE scheme called TFHE, or FHE over the Torus \cite{tfhe}. Boolean FHE is a form of FHE that supports bitwise logical operations such as bitwise union and intersection. By deconstructing programming logic and conditions into binary circuits and Boolean arithemtic, we are able to create alternative representations of logical operations that do not rely on evaluations of necessary conditions, thereby enabling \textit{blind} programming. In the upcoming sections, we describe in detail how Boolean FHE and its Boolean properties are used to ultimately enable encrypted general programming without IRDE.



\section{Related Work} \label{sec:related}
\subsection{Privacy-Preserving Machine Learning}\label{rw-ppml}
There have been much progress made in the field of PPML and FHE in the last twenty years. Prior works \cite{kim2018logistic,kim2018secure} constructed privacy-preserving logistic regression models using FHE with polynomial approximation of activation functions. The practice of finding polynomial approximations of functions extends beyond activation functions and is commonly used in FHE to allow for faster computation \cite{panda2022polynomial,dang2024accurate,lee2023precise}. Cryptonets \cite{gilad2016cryptonets} opened the door for neural network evaluation by successfully adopting FHE into the inference process of an already trained neural network. Later, Hesamifard et al. introduced CryptoDL, a framework that allows training and evaluation of deep neural networks with client-server interaction for noise reduction \cite{hesamifard2018privacy}. Using TFHE (or FHE over the torus) \cite{tfhe}, results indicating the feasibility of deep neural network inference with FHE were shown \cite{cryptoeprint:2021/091} and later for tree-based inference in \cite{frery2023privacy}. Following works \cite{cryptoeprint:2021/091,frery2023privacy,stoian2023deep} use \textit{programmable bootstrapping} as part of their FHE scheme as opposed to \textit{leveled mode} used by \cite{kim2018logistic,kim2018secure,gilad2016cryptonets}. Using FHE with \textit{leveled mode} allows for faster computation compared to \textit{bootstrapped mode} but can only compute a predetermined number of products due to noise growth. This poses a problem with respect to scalability of models. On the other hand, FHE with \textit{bootsrapped mode} enables noise reduction whenever noise reaches a certain threshold, allowing for evaluation of complex circuits and functions. Our framework uses \textit{programmable bootstrapping} to allow scalable and non-interactive computation from the server. 

\subsection{Privacy-Preserving Decision Trees}
\subsubsection{Multi-Party Computation Approaches}
Concerning decision trees, the first attempt at privacy-preserving decision tree training involved Secure Multi-Party Computation (SMPC) \cite{lindell2000}. Prior works in \cite{de2014practical,du2002building} also considered SMPC techniques for decision tree learning. A method for training and inferring decision trees with a twin-cloud architecture using additive secret sharing was also explored in \cite{twin-cloud}. These approaches generally involve two or more parties adhering to strict security and privacy standards and can be made vulnerable if one or more participating parties collude with each other or break their security parameters. In contrast, a non-interactive client-server model used in our work simplifies the architecture and allows the server to handle encrypted computations without any interaction from the client who holds the private-key.

\subsubsection{Client-Server Models}
In the context of a simple client-server model where privacy-preserving computations with FHE are conducted on the server, following prior works \cite{cong2022sortinghat,mahdavi2023level,akavia2022privacy} demonstrated non-interactive decision tree inference while \cite{sun2018private,alex2022private} relied on interactive comparison protocols. However, non-interactive training of decision trees has yet to be achieved with the state-of-the-art IRDE performance requiring d-rounds of IRDE for tree-depth of d \cite{akavia2022privacy}. The seemingly infeasible nature of non-interactive training of decision trees with FHE is one that is prevalent in privacy-preserving machine learning as a whole. In neural networks for example, while non-interactive, encrypted inference with FHE has seen advancements \cite{gilad2016cryptonets,cryptoeprint:2021/091}, non-interactive training has yet to be achieved much like with decision trees.
\\
Table \ref{tab:tableSummaryPriorWorks} summarizes prior works based on their ability or inability to perform non-interactive training and inference of encrypted decision tree modeling. Prior works in \cite{de2014practical,du2002building,int_training_1,lindell2000,int_training_2,akavia2022privacy} performed privacy-preserving training with IRDE while many existing works ignored training altogether and focused on encrypted inference \cite{int_pred_1,bost2014machine,int_pred_2,int_pred_3,int_pred_4,cong2022sortinghat,mahdavi2023level,lu2018non,tueno2020non}. In a departure from existing works, our work presents a programming framework that enables both encrypted training and inference of PPML models without IRDE. We do this by programmatically enabling \textit{blind} executions of control structures and other non-arithmetic operations, correctly and effectively computing such operations without evaluating their conditions or results. We demonstrate our framework using privacy-preserving decision trees showcasing encrypted training and inference while eliminating IRDE entirely. To the best of our knowledge, this is the first time both training and inference of privacy-preserving decision trees with FHE were modeled without IRDE.

\begin{table}[htbp]
    \centering
    \caption{Table Summary of Prior Works}
    \label{tab:tableSummaryPriorWorks}
    \begin{tabularx}{\columnwidth}{|X|X|X|}
        \hline
        \textbf{Approaches} & \textbf{Non-Interactive Training} & \textbf{Non-Interactive Inference} \\
        \hline
        \cite{de2014practical,du2002building,int_training_1,lindell2000,int_training_2} & \ding{55} & N/A \\ 
        \hline
        \cite{int_pred_1,bost2014machine,int_pred_2,int_pred_3,int_pred_4} & N/A & \ding{55} \\ 
        \hline
        \cite{cong2022sortinghat,mahdavi2023level,lu2018non,tueno2020non} & N/A & \checkmark \\ 
        \hline
        \cite{akavia2022privacy} & \ding{55} & \checkmark \\ 
        \hline
        \textbf{BEF} & \checkmark & \checkmark \\
        \hline
    \end{tabularx}
\end{table}

\section{Blind Evaluation Framework (BEF)}\label{sec:bef}
Programming complex logic and control flow with FHE presents challenges and typically require client-assisted computation. For example, one analysis notes that high-level of communication (IRDE) required for comparison operations and conditional operations are among the biggest reasons making FHE unsuitable in machine learning algorithms \cite{chialva2018conditionals}. As a result, non-interactive PPML has not seen serious progress particularly when it comes to training. To surmount this barrier, we introduce the Blind Evaluation Framework (BEF), a novel programming framework that facilitates \textit{blind} executions of complex logical programs in the encrypted domain of FHE without IRDE. 
\\
The essence of BEF lies in deconstructing programming logic into binary circuits and Boolean arithmetic in ways that enable correct executions of logical operations even when the necessary conditions are unknown and cannot be evaluated due to encryption. BEF is constructed from a series of foundational algorithms that work in conjunction to perform 'blind' evaluations of expressions, conditions, and control flow of programs. As detailed in the following subsections, these algorithms include \textit{blind comparison}, \textit{blind selection}, and \textit{blind ordering}. Together, they make up the building blocks of BEF and are crucial for implementing more complex operations. They enable a range of functionalities, from basic conditional branching based on encrypted comparison results to more advanced applications like argmax and argmin calculations, and even sorting operations, all without requiring knowledge of the comparison results or conditional information.
\\
This section focuses on illustrating these fundamental algorithms through simple, yet demonstrative examples. The motivation behind this approach is twofold. Firstly, by showcasing these algorithms with straightforward and simplistic examples, this paper aims to underscore their adaptability and efficacy in a variety of contexts. Secondly, these examples serve as a proof of concept, highlighting how such basic building blocks can be extended and integrated into more complex and powerful operations. These foundational concepts address fundamental constraints in FHE and pave the way for robust and powerful applications in encrypted computing with FHE. This foundational approach is particularly pertinent in the construction of Privacy-Preserving Machine Learning (PPML) models, where these algorithms can be utilized for both training and prediction without IRDE.

\subsection{Blind Comparison}
\begin{figure}[htbp]
  \centering
  \includegraphics[width=\columnwidth]{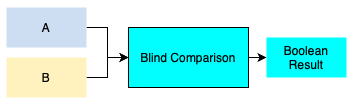}
  \caption{High-level diagram of blind comparison}
  \label{fig:diagram-comp}
\end{figure}

Blind comparison performs a magnitude comparison on encrypted inputs and returns an encrypted result. Figure \ref{fig:diagram-comp} shows a high-level, input-output diagram of the blind comparison operation. Note that inputs \textit{A} and \textit{B} are encrypted with FHE as is the output, \textit{Boolean Result}.
\\
Blind comparison circuits can be constructed with bitwise logic, using Boolean FHE schemas, as demonstrated below in Figure \ref{fig:twobitcircuit}. Using the circuit, a more detailed representation of Figure \ref{fig:diagram-comp} is shown in Algorithm \ref{algo:blindcomparison}. Here, the function processes two 2-bit inputs, $A$ and $B$, encrypted with FHE and returns their encrypted comparison results. The output is a tuple of encrypted values, indicating the comparison outcome without revealing any actual data. If $A > B$ holds true, then the tuple will return $(\mathcal{E}(1), \mathcal{E}(0))$, and $(\mathcal{E}(0), \mathcal{E}(1))$ otherwise, where $\mathcal{E}(\cdot)$ denotes that $\cdot$ is encrypted. This process can be modified to support greater-than-equal or less-than-equal operations. The simplicity of blind comaprison is deceptive as it forms a cornerstone for executing more complex control structures with BEF.
\\
In prior works, the limiting factor in comparison operations with FHE was the lack of usability of encrypted comparison results. Hence, previous works that performed encrypted comparisons used IRDE \cite{kiss2019sok, bost2014machine,sun2018private} or used encrypted comparison results in limited contexts within machine learning that didn't involve training \cite{cong2022sortinghat,mahdavi2023level,frery2023privacy}. However, with BEF, we present ways to take encrypted comparison results and use them to perform \textit{blind}, but correct, logical programming without IRDE. The following sections demonstrate how blind programming is achieved using encrypted conditions such as encrypted comparison results.

\begin{figure}
  \centering
  \includegraphics[width=\columnwidth]{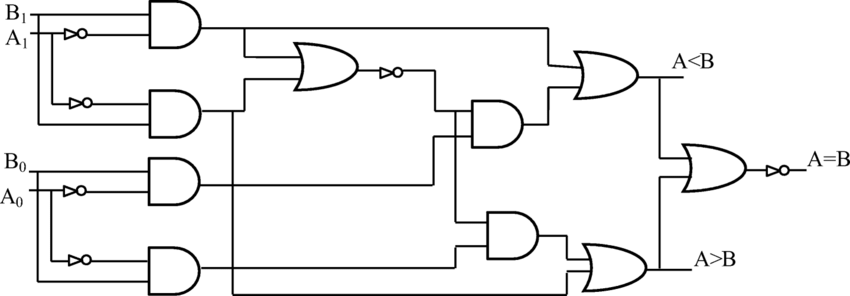}
  \caption{2-bit magnitude comparison circuit \cite{comparatorcircuit}}
  \label{fig:twobitcircuit}
\end{figure}

\begin{algorithm}[H]
    \caption{Blind Comparison}
    \label{algo:blindcomparison}
    \begin{algorithmic}[1]
        \STATE $A = \{a_0, a_1\}$
        \STATE $B = \{b_0, b_1\}$
        \STATE \textbf{gt} is 1 if $A > B$, 0 otherwise
        \STATE \textbf{lte} is 1 if $A <= B$, 0 otherwise
        \STATE
        \STATE {\textsc{BlindComparison}}$(\mathcal{E}(A)$, $\mathcal{E}(B))$
            \STATE \hspace{0.5cm} $\mathcal{E}(gt) \gets \mathcal{E}(a_1) \land \neg \mathcal{E}(b_1) \lor \mathcal{E}(a_0) \land \neg \mathcal{E}(b_1) \land \neg \mathcal{E}(b_0)$    \\
            \hspace{0.5cm}  $\lor \mathcal{E}(a_1) \land \mathcal{E}(a_0) \land \neg \mathcal{E}(b_0)$
            \STATE
            \STATE \hspace{0.5cm} $\mathcal{E}(lt) \gets \neg \mathcal{E}(a_1) \land \mathcal{E}(b_1) \lor \neg \mathcal{E}(a_0) \land \mathcal{E}(b_1) \land \mathcal{E}(b_0)$ \\
            \hspace{0.5cm} $\lor \neg \mathcal{E}(a_1) \land \neg \mathcal{E}(a_0) \land \mathcal{E}(b_0)$
            \STATE
            \STATE \hspace{0.5cm} $\mathcal{E}(eq) \gets \neg (\mathcal{E}(gt) \lor \mathcal{E}(lt))$
            \STATE
            \STATE \hspace{0.5cm} $\mathcal{E}(lte) \gets \mathcal{E}(lt) \lor \mathcal{E}(eq)$
            \STATE
            \STATE \hspace{0.5cm} \textbf{return} $\mathcal{E}(gt), \mathcal{E}(lte)$
    \end{algorithmic}
\end{algorithm}

\subsection{Blind Selection}
\begin{figure}
  \centering
  \includegraphics[width=\columnwidth]{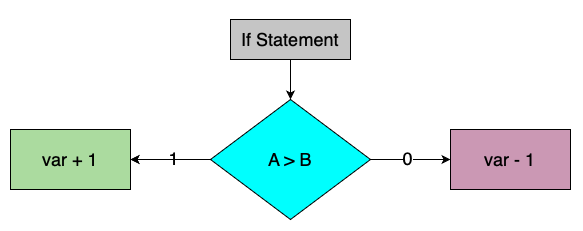}
  \caption{If-else statement diagram}
  \label{fig:selection_one}
\end{figure}

\begin{figure}
  \centering
  \includegraphics[width=\columnwidth]{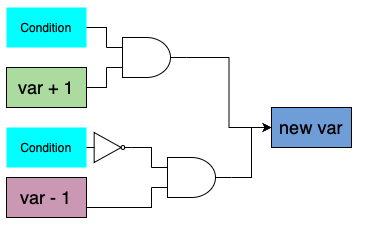}
  \caption{Blind selection using binary circuit}
  \label{fig:selection_two}
\end{figure}

A major challenge with control structures in FHE is the inability to perform conditional branching as is typically done in conventional programming, such as in Algorithm \ref{algo:condBranching}. A diagram representation of the same branching operation is shown in Figure \ref{fig:selection_one}. This limitation arises because condition variables (shown in cyan in Figure \ref{fig:selection_one}), such as those derived from blind comparisons, remain encrypted and their true values are thus indiscernible. Indeed, attempting to use encrypted data or variables as conditions in traditional \textit{if-else} statements results in a compile-time error in FHE. 
\\
To solve this challenge, we present blind selection, a methodology that enables selection of encrypted value, branch, function, etc. based on an encrypted condition. Blind selection works by deconstructing conditional logic into Boolean arithmetic, whereby branches are preserved or discarded based on a given encrypted condition. This is achieved using bitwise intersections (AND operations in digital logic), simulating the selection of the correct branch while effectively collapsing or nullifying the others, as shown in Algorithm \ref{algo:blindselection}. A diagram representation of blind selection is shown in Figure \ref{fig:selection_two}, demonstrating how blind selection achieves the same, correct result as in Figure \ref{fig:selection_one} without knowledge of the condition variable (in cyan), and thus remaining compatible with FHE. A version closely resembling blind selection is introduced in \cite{chialva2018conditionals}, with a \textit{collapse-or-preserve} methodology achieving essentially the same results as blind selection from BEF. However, this method was limited to selection of numerical variables in leveled-mode while our blind selection applies to any \textit{path choosing} operation such as executing different functions based on given encrypted conditions.
\\
It is worth noting that blind selection is not confined solely to conditional statements involving blind comparison results. Rather, it can be used any time a branching operation hinges on a Boolean condition variable.

\begin{algorithm}[htbp]
    \caption{Example of Conventional Conditional Branching}
    \label{algo:condBranching}
    \begin{algorithmic}[1]
        \STATE \textbf{Input:} A, B, var
        \STATE \textbf{Output:} var modified based on comparison result
        \STATE
        \IF{A $>$ B}
            \STATE var $\gets$ var + 1
        \ELSE
            \STATE var $\gets$ var - 1
        \ENDIF

    \end{algorithmic}
\end{algorithm}

\begin{algorithm}[htbp]
    \caption{Blind Selection}
    \label{algo:blindselection}
    \begin{algorithmic}[1]
        \STATE \textbf{Input:} $\mathcal{E}(A), \mathcal{E}(B), \mathcal{E}(var)$
        \STATE \textbf{Output:} $\mathcal{E}(new\_var)$
        \STATE
        \STATE {\textsc{BlindSelection}}$(\mathcal{E}(A), \mathcal{E}(B))$
        \STATE \hspace{0.5cm} $(\mathcal{E}(gt), \mathcal{E}(lte)) = {\textsc{blindcomparison}}{\mathcal{E}(A), \mathcal{E}(B)}$
        \STATE \hspace{0.5cm} $\mathcal{E}(new\_var)$ $\gets$ $\mathcal{E}(gt) * (\mathcal{E}(var) + 1)$ \\
        \hspace{0.5cm} + $\mathcal{E}(lte) * (\mathcal{E}(var) - 1)$
        \STATE \hspace{0.5cm} \textbf{return} $\mathcal{E}(new\_var)$
    \end{algorithmic}
\end{algorithm}

\subsection{Blind Ordering}
\begin{figure}
  \centering
  \includegraphics[width=\columnwidth]{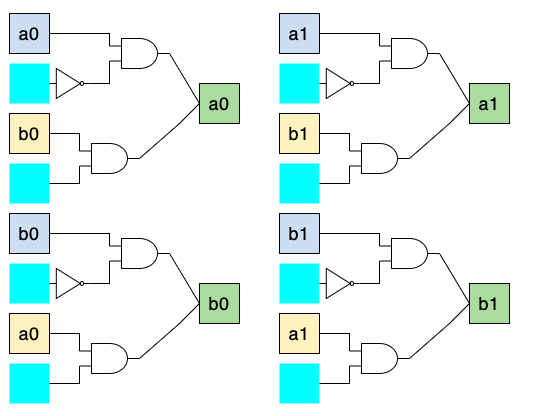}
  \caption{Diagram representation of blind ordering}
  \label{fig:ordering}
\end{figure}

Blind ordering serves as a key building block for advanced operations such as argmin/max calculations and sorting. This process involves swapping a pair of values, either in ascending or descending order, using blind comparison as the swapping condition and blind selection to facilitate the swapping process. Blind ordering is used to ensure the desired order of values in a list of unknown (encrypted) values. By ensuring the correct order of values in a given tuple or list, blind ordering can facilitate the selection of the desired value within a pair without knowledge of their actual content.
\\
A diagram representation of blind ordering of tuple $[A,B]$ where inputs $A$ and $B$ are two-bit binary strings ($A = \{a0,a1\}$, $B = \{b0, b1\}$) is shown in Figure \ref{fig:ordering}. Using the blind comparison result (in cyan) from Figure \ref{fig:diagram-comp}, blind selection is performed to either preserve $a0$ or replace it with $b0$ depending on the blind comparison result, and vice versa. This process is performed bit-wise for inputs $A$ and $B$. By taking advantage of blind comparison and blind selection, a tuple of unknown and encrypted values can be arranged in ascending or descending order without having to evaluate or gain knowledge of the values directly.
\\
The strength of blind ordering lies in its ability to make informed selections in an encrypted environment. This capability makes blind ordering an invaluable tool for implementing blind selection methods, enabling the accurate and secure identification of desired values across a broad spectrum of applications in encrypted programming with FHE.

\begin{algorithm}
    \caption{blind ordering}
    \label{algo:swapping}
    \begin{algorithmic}[1]
        \STATE \textbf{Output from blind comparison:} $\mathcal{E}(gt)$, $\mathcal{E}(lt)$
        \STATE \textbf{Input:} array[$\mathcal{E}(A)$, $\mathcal{E}(B)$] where both A and B are n-bit binary strings
        \STATE \textbf{Output:} array[$\mathcal{E}(A)$, $\mathcal{E}(B)$] or array[$\mathcal{E}(B)$, $\mathcal{E}(A)$]
        \STATE
        \STATE {\textsc{blindOrdering}}$(array)$
            \STATE $\mathcal{E}(gt)$, $\mathcal{E}(lt)$ = \textsc{blindcomparison}$(\mathcal{E}(A)$, $\mathcal{E}(B))$
            \FOR{$i \gets 0$ to $n$}
                \STATE $array[0][i] = (\mathcal{E}(lt) \land array[1][i])$ \\
                \hspace{0.5cm} $\lor (\mathcal{E}(gt) \land array[0][i])$
                \STATE $array[1][i] = (\mathcal{E}(lt) \land array[0][i])$ \\
                \hspace{0.5cm} $\lor (\mathcal{E}(gt) \land array[1][i])$
            \ENDFOR
        \STATE \textbf{return} $array$
    \end{algorithmic}
\end{algorithm}

\subsection{Applications of BEF}
\begin{figure}
  \centering
  \includegraphics[width=\columnwidth]{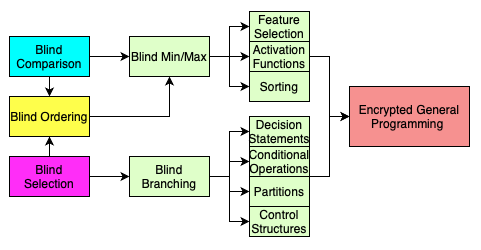}
  \caption{Applications of BEF}
  \label{fig:applications}
\end{figure}

With blind comparison, selection, and ordering making up the building blocks of BEF, more complex algorithms and programming structures can be constructed to ultimately enable encrypted general programming as shown in Figure \ref{fig:applications}. For example, blind ordering can be iterated over a list to perform argmin or argmax operations. When this operation is nested inside another iterator, blind sorting can be achieved as shown in Algorithm \ref{algo:sorting}.
\\
Using BEF and its many blind algorithms and applications, various PPML models can be constructed that do not rely on IRDE. For example, encrypted K-Nearest Neighbors (KNN) typically required communication protocols to resolve comparisons and other operations or allow significant precision loss by approximating challenging operations \cite{wang2023outsourced,behera2022preserving}. With BEF however, KNN can be performed in encrypted space without evaluating the actual distances of neighbors using blind comparisons and blind selection, achieving privacy-preserving KNN without IRDE or precision loss. In prior works, without techniques like blind selection, encrypted distance calculations could be achieved but the actual selection of the appropriate data points required communication (IRDE) or multi-party protocols. Likewise, threshold operations (ReLU, MaxPooling, etc.) often required polynomial approximations at the cost of precision loss and limited input range \cite{dang2024accurate} or communication operations (IRDE) as in \cite{nguyen2023hefun}. One reason why such compromises were necessary in the past was the embedded comparison operation inside any thresholding operation. Encrypted comparison results are useless without a paradigm that can program blindly without evaluation. However, with BEF, blind selection can be used to seamlessly integrate these threshold operations without precision loss or the need for IRDE. For a comprehensive demonstration, we choose encrypted decision tree training and prediction without client-assisted computation or IRDE to showcase how BEF is applied in PPML from start to finish. In addition, the blind algorithms and applications mentioned in this work is by no means an exhaustive list; readers are encouraged to create additional blind algorithms and use-cases in PPML to further advance encrypted general programming that do not rely on IRDE. 
\\
The building blocks of BEF and some demonstrative applications such as blind sorting are available in \cite{myCode}. While the computational overhead introduced by FHE currently poses challenges for the practical implementation of the Blind Evaluation Framework (BEF), it is important to recognize the need for a framework that can perform complex computations like machine learning training using FHE without IRDE or other compromises.
As the field of FHE continues to evolve, with ongoing advancements in computational efficiency, there is a strong potential for BEF to become more practically viable in the future. This progression promises to unlock the full capabilities of BEF, making it an integral part of privacy-preserving computational solutions.

\begin{algorithm}[htbp]
    \caption{Blind Sorting}
    \label{algo:sorting}
    \begin{algorithmic}[1]
        \STATE {\textsc{blindSorting}}$(array)$
            \FOR{$i$ in 0..array.length}
                \STATE $j = i$
                \WHILE{$j > 0$}
                    \STATE {\textsc{blindSwapping}}$(array[j-1], array[j])$
                    \STATE $j = j - 1$
                \ENDWHILE
            \ENDFOR
    \end{algorithmic}
\end{algorithm}

\subsection{Challenges of Decision Tree Training and Prediction with FHE}
Implementing decision tree training and prediction using Fully Homomorphic Encryption (FHE) without Blind Evaluation Framework (BEF) presents a unique set of challenges. These challenges stem from the fundamental characteristics of FHE and the intricate requirements of decision tree algorithms.
\\
One major challenge in training decision trees under FHE is handling recursive data partitioning. Decision trees rely on evaluating input features and corresponding labels to create branches and nodes, a process inherently difficult with FHE. FHE encrypts data so that comparisons or evaluations, necessary for partitioning in decision trees, are not easily achievable. This limitation significantly hinders the ability to effectively split data at each node, a crucial step in the tree's construction. Moreover, the prediction phase in decision trees involves traversing the tree and making decisions at each node based on encrypted input. This process requires evaluating conditions and choosing the appropriate path down the tree to reach a leaf node. In the context of FHE, where direct evaluation of conditions is not feasible, executing this traversal becomes a complex task. 
\\
BEF addresses these challenges by enabling key operations like encrypted data comparisons and control structure executions. Using BEF, training and prediction of machine learning models with FHE can be made feasible without IRDE.

\begin{algorithm}[htbp]
    \caption{Decision Tree Training}
    \label{algo:decision_tree_training}
    \begin{algorithmic}[1] 
        \STATE \textbf{X:} feature data where $X^i$ refers to column data
        \STATE \textbf{y:} class labels 
        \STATE {\textsc{TrainDecisionTree}}$(X, y)$
            \IF{all X belong to the same class} 
                \STATE \textbf{return} a leaf node with that class label
            \ENDIF
            
            \IF{$X^i$ is empty}
                \STATE \textbf{return} a leaf node with the majority class label among remaining examples
            \ENDIF
            
            \STATE Choose the $X^i$ $i$ that best splits the examples
            \STATE Create a new decision tree node $N$ with attribute $i$
            
            \FORALL{possible value $V$ of attribute $i$}
                \STATE $X_v \gets$ subset of examples where attribute $i$ has value $V$
                \IF{$X_v$ is empty}
                    \STATE Attach a leaf node with the majority class label among remaining examples to node $N$
                \ELSE
                    \STATE Attach the subtree \textsc{TrainDecisionTree}($X_v$, $X^i - i$) to node $N$
                \ENDIF
            \ENDFOR
            
            \STATE \textbf{return} the root node $N$ of the decision tree
    \end{algorithmic}
\end{algorithm}

\begin{algorithm}[htbp]
    \caption{Decision Tree Evaluation}
    \label{algo:decision_tree_eval}
    \begin{algorithmic}
        \STATE \textbf{x:} input data to be evaluated
        \STATE {\textsc{Prediction}}$(node,x)$
            \IF{node is leaf node}
                \STATE \textbf{return} $node.class-label$
            \ENDIF
            
            \IF{$x.value < node.threshold$}
                \STATE \textbf{return} {\textsc{Prediction}}$(node.left, x)$
            \ELSE
                \STATE \textbf{return} {\textsc{Prediction}}$(node.right, x)$
            \ENDIF
    \end{algorithmic}
\end{algorithm}

\section{Privacy-Preserving Decision Tree Training using Blind Evaluation Framework (BEF)}\label{sec:ppdtt}
Using BEF, a privacy-preserving decision tree using FHE can be trained and used for prediction without client-server communication protocols (IRDE) while retaining a simple client-server model. Unlike client-server models that require communication protocols with IRDE, this client-server model using BEF utilizes a \textit{send-and-forget} model, where the client can be offline and free from computational commitment as soon as encrypted data are sent to the server.
\\
Implementing encrypted decision tree training using BEF requires designing programmatic control structures in a way that does not rely on evaluation and knowledge of any necessary condition or expressions. Moreoever, control structures have to be designed to execute correctly despite having no knowledge of what branch of the program it is currently on or what the control flow of the program is. Designing and implementing such programs are challenging and require a departure from the way standard programs are written. Therefore, it is helpful to divide the training process into high-level steps. 
\\
The first step is the feature selection phase, where the feature (decision node) to be used as the root node is selected from the original dataset. Second, with the ideal feature selected as the root node, the data splitting phase is executed in order to proceed to the left and right child nodes of the root node. Next, the tree growing phase is initiated where the decision tree continues to recursively build and train using aspects of the feature selection phase and data splitting phase. Lastly, the training procedure is stopped and results returned to the client during termination phase when a chosen termination criteria are met. For simplicity, this paper assumes the dataset consists of binary categorical data and labels. However, this framework is compatible with any other data type (floating point numbers, integers, etc.) as long as they can be represented in binary format.

\subsection{Feature Selection Phase}
In decision tree training, feature selection is a critical phase, where the goal is to identify the feature that minimizes expected classification errors. Various feature scoring algorithms for feature selection such as the Gini index and Information Gain, aid in this process. The Gini index evaluates the likelihood of incorrect classifications for each feature, known as Gini impurity, aiming to select the feature with the lowest impurity. Information Gain, in contrast, measures the usefulness of each feature, selecting the one that offers the highest gain. While these feature scoring algorithms are crucial, they are mostly made up of straight-forward mathematical operations and their implementations with FHE have been previously established \cite{akavia2022privacy} and are not the primary focus of our work. Instead, our contribution lies in our ability to blindly select the correct feature using encrypted feature importance scores without IRDE, which has not been shown in previous works. Using blind sorting, which leverages blind comparison and blind swapping, the list of features can be sorted in ascending order based on their encrypted feature importance values. Blind sorting ensures that the first element in the blindly sorted list will be the feature with the lowest Gini impurity (if using information gain, the last element will be the desired feature). Leveraging BEF, a decision node can be selected for the root node of the decision tree accurately and without evaluating information about any of the features. Algorithm \ref{algo:FSP} gives a high-level outline for this phase.

\begin{algorithm}
    \caption{Feature Selection Phase}
    \label{algo:FSP}
    \begin{algorithmic}[1]
        \STATE {\textsc{FeatureSelection}}$(dataset)$
            \FOR{features in dataset}
                \STATE $gini_i\gets giniImpurity(feature_i)$
                \STATE $tupleList \gets (feature_i.column, gini_i)$
            \ENDFOR
            \STATE {\textsc{blindSorting}}$(tupleList)$
            \STATE \textit{remove the feature corresponding to tupleList[0] from the dataset}
            \STATE \textbf{return} tupleList[0]
    \end{algorithmic}
\end{algorithm}

\subsection{Data Partitioning Phase}
In decision tree training, once a feature is selected for the root node, the dataset needs to be divided into subsets for the tree's left and right branches. However, with FHE, direct partitioning of the dataset based on encrypted comparison results is not feasible.
\\
To navigate this challenge, BEF can be used to achieve \textit{soft-partitioning}. This technique simulates the splitting of data without actual partitioning of the dataset. It involves the creation of a condition vector, which is derived from the feature data calculations of the root node. This vector is then used as the condition for blind selection to guide the distribution of data to the left and right branches of the tree.
\\
In practice, for each entry in the condition vector, its value determines the branch to which the corresponding row of the dataset belongs. For instance, an entry marked as encrypted false implies that the row belongs to the left branch, while encrypted true indicates belonging to the right branch.
\\
This approach, as demonstrated in Algorithm \ref{algo:DPP}, lies in passing the entire original dataset, along with the condition vector, to both branches. This method leverages blind selection to facilitate accurate feature selection calculations for both branches. As a result, each node at the same level of the tree contains identical data encrypted with FHE, yet can still compute the necessary operations accurately using the condition vector.
\\
A key advantage of this method in addition to not requiring IRDE is its ability to maintain model privacy. By employing soft-partitioning with blind selection, the server can perform necessary data splitting operations without needing to decrypt the data. Moreover, this approach conceals any information about the subsets' sizes and imbalances, as well as the overall structure of the tree, since even empty nodes contain the same data as other nodes at the same level. Thus, BEF effectively enables data partitioning in encrypted decision tree training without IRDE, preserving both data and model security.

\begin{algorithm}
    \caption{Data Partitioning Phase}
    \label{algo:DPP}
    \begin{algorithmic}
        \STATE {\textsc{DataPartitioning}}$(dataset, node)$
            \STATE $selectedFeature \gets {\textsc{FeatureSelection}}(dataset)$
            \STATE $node.threshold \gets$ selected threshold for this feature
            \STATE \textbf{$vector_{condition}$} $\gets selectedFeature.column$
            \STATE $node.left \gets (dataset, vector_{condition})$
            \STATE $node.right \gets (dataset, vector_{condition})$
    \end{algorithmic}
\end{algorithm}

\subsection{Tree Growing Phase}
With a feature selected for the root node and data partitioning phase having been executed, the next step is to continue growing the tree by populating child nodes and branching out until some termination criterion is met. Gini calculations on the child nodes are calculated much the same as the root node, guided by blind selection and the condition vector to perform appropriate computations for either branches. Data partitioning phase and feature selection phase that are slightly modified to support subset calculations are recursively performed for each child node at each level of the decision tree.

\begin{algorithm}
    \caption{Tree Growing Phase}
    \label{algo:TGP}
    \begin{algorithmic}
        \STATE {\textsc{TreeGrowing}}$(dataset, vector_{condition}, node)$
            \IF{termination criteria is met}
                \STATE \textbf{return} root
            \ENDIF
            \STATE perform feature selection and data partitioning phases
            \STATE {\textsc{TreeGrowing}}$(dataset, vector_{condition}, node.left)$
            \STATE \textsc{TreeGrowing}$(dataset, vector_{condition}, node.right)$
    \end{algorithmic}
\end{algorithm}

\subsection{Termination Phase}
Decision trees are prone to overfitting and can overfit perfectly to training data unless some stopping criterion is employed to halt the training process. In the case of privacy-preserving decision tree paradigm proposed in this work, throughout the recursive process of tree growing phase, valid data entries (as determined by the condition vector) for each node in successive levels will grow smaller until every node in a given level no longer has any valid data left. Because of the nature of FHE, the training process will be blind to the fact that there is no valid data remaining to be trained and that the tree expansion should be stopped. Therefore, stopping mechanism in the proposed FHE decision tree is necessary not only to minimize overfitting, but also to stop the tree from training indefinitely.
There are many stopping criteria that can be employed to stop the training of decision trees. One of the most effective and straightforward methods is to simply set a maximum depth for a tree so that the tree stops training at a given depth even if further training can be done on remaining data. This method minimizes overfitting and can be implemented in a straightforward manner in the proposed FHE decision tree. An integer parameter denoting the maximum depth for the tree can be set by the client and passed to the server in plaintext to control the number of tree growing iterations performed during training. While providing this information in plaintext may appear to leak information about the tree, this information would already be known to the server if it was to perform inference with the decision tree, and therefore the solution proposed in this paper does not leak any information beyond what would already be known by the server.

\subsection{Integrating the Phases}
Having defined the different phases of the privacy-preserving decision tree training paradigm, they can be integrated into one unifying algorithm to demonstrate the proposed paradigm in full, as shown in Algorithm \ref{algo:fullTraining}. 

\begin{algorithm}
    \caption{Privacy-Preserving Paradigm for Decision Tree Training}
    \label{algo:fullTraining}
    \begin{algorithmic}
        \STATE \textbf{root:} starting node, entry point for decision tree
        \STATE \textsc{TrainingAlgorithm}$(dataset)$
            \FOR{features in dataset}
                \STATE $gini_i \gets giniImpurity(feature_i)$
                \STATE $tupleList \gets (feature_i.column, gini_i)$
            \ENDFOR
            \STATE \textsc{blindSorting}$(tupleList)$
            \STATE \textbf{selectedFeature} $\gets tupleList[0]$
            \STATE \textit{remove the feature corresponding to tupleList[0] from the dataset}

            \STATE $root.threshold \gets selected threshold for this feature$
            \STATE $vector_{condition}\gets selectedFeature.column$

            \IF{termination criteria is met}
                \STATE \textbf{return} root
            \ENDIF
            \STATE \textsc{TreeGrowing}$(dataset, vector_{condition}, root.left)$
            \STATE \textsc{TreeGrowing}$(dataset, vector_{condition}, root.right)$
    \end{algorithmic}
\end{algorithm}

\section{Privacy-Preserving Decision Tree Prediction with Blind Evaluation Framework}\label{sec:ppdtp}
With the trained tree on the server, inference can be performed to evaluate new, incoming data. Although inference is not the focal point of this paper and indeed many papers in the past have demonstrated decision tree inference with FHE, discussing the inference process is necessary to showcase how the decision tree trained using BEF can be seamlessly deployed for prediction. The prediction protocol takes a similar approach taken by \cite{akavia2022privacy} and converts the trained tree into a chain of polynomial equations. 
\\
For prediction over binary, categorical data $x^i \in X$, taking the right branch is correct if $x^i$ is $\mathcal{E}(True)$ and taking the left branch is correct if $x^i$ is $\mathcal{E}(False)$. Therefore, line 5 in Algorithm \ref{algo:pred_binary} ensures that the correct path of the tree is selected through blind selection.
\\
For prediction in cases where it is necessary to have continuous or non-binary threshold values, blind comparison from Algorithm \ref{algo:blindcomparison} would need to be employed to compare the appropriate feature value from input data with the threshold value at any given node of the tree. This process is shown in Algorithm \ref{algo:pred_general} where the results of blind comparisons are used for blind selection of the correct path in the tree.  
\\
This methodology of converting the trained decision tree into a polynomial equation representation and using blind selection ensures that correct result is obtained in prediction while maintaining privacy of data and even ensuring that the correct path is hidden to the server during runtime. This prediction protocol is independent of  training protocol and can be used for any decision tree, plaintext or encrypted.

\begin{algorithm}
    \caption{Prediction Protocol: Binary}
    \label{algo:pred_binary}
    \begin{algorithmic}[1]
        \STATE \textbf{predict($node$, $x$):} where $node$ refers to an FHE decision tree node containing feature information and $x$ denoting the incoming validation data
        
        \IF{$node$ contains leaf value}
            \STATE \textbf{return} $node.leaf\_value$
        \ELSE
            \STATE \textbf{return} $x^{node} \cdot predict(node.right, x)$ + \\ $x^{node} \cdot predict(node.left, x)$
        \ENDIF
    \end{algorithmic}
\end{algorithm}

\begin{algorithm}
    \caption{Prediction Protocol:General}
    \label{algo:pred_general}
    \begin{algorithmic}[1]
        \STATE In the case where node and input are not binary, categorical data, requires threshold comparison to determine the correct evaluation path

        \IF{$node$ contains leaf value}
            \STATE \textbf{return} $node.leaf\_value$
        \ELSE
            \STATE \textbf{return} $\textsc{blindcomparison}(x^{node}, node.threshold) \cdot predict(node.right, x)$ + \\
            $\textsc{blindcomparison}(x^{node}, node.threshold) \cdot predict(node.left, x)$
        \ENDIF
        
    \end{algorithmic}
\end{algorithm}

\section{Evaluations and Results}
This section goes over the analysis and results of BEF algorithms as well as privacy-preserving, non-interactive decision tree training and prediction with BEF. All experiments were conducted using AMD EPYC 7763 processor with 128 cores. For implementation, a TFHE \cite{tfhe} variant for the Rust programming language called TFHE-rs \cite{TFHE-rs} was used with details of TFHE-rs available in \cite{chillotti2021programmable}. 

\subsection{BEF Evaluations}
\begin{figure}
  \centering
  \includegraphics[width=\columnwidth]{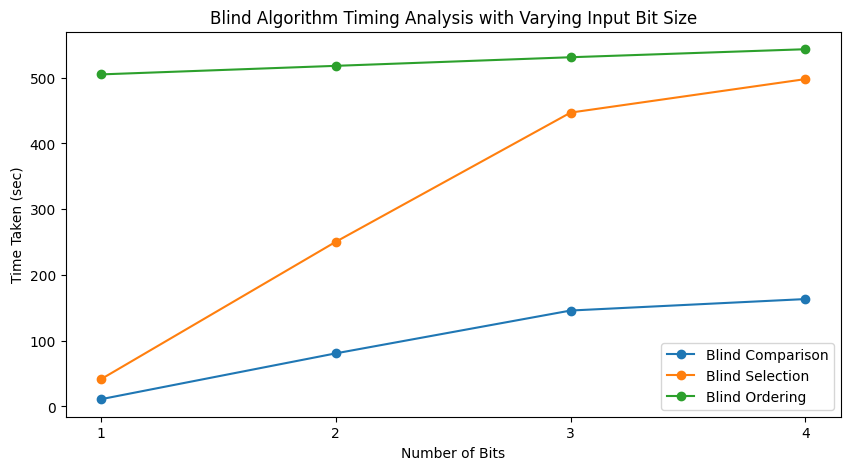}
  \caption{Timing of blind operations with varying input size (as number of bits)}
  \label{fig:bitwise-blindalgo}
\end{figure}

This section goes over runtime analysis of the building blocks of BEF. Table \ref{tab:blind-runtime} shows the runtime complexity of various blind operations with $m$ corresponding to the number of input bits. The actual execution times of the building blocks of BEF with varying input size in the number of bits is shown in Figure \ref{fig:bitwise-blindalgo}. While blind comparison, selection, and ordering all theoretically have linear complexity, the actual execution timings did not show consistent linear growth in runtime. One explanation for this discrepancy is that the overall computational slowness of FHE may overshadow the runtime complexity of bit-wise operations and could lead to high growth in runtime initially and subsequent slower growth in runtime with increasing input bits. With $x$ denoting the number of bits used as input, Equations in \ref{eq:bitwise-blindalgo} extrapolate the execution times (in seconds) of blind comparison (as $f(x)$), blind selection (as $g(x)$), and blind ordering (as $h(x)$) as polynomial equations. Putting them together to reflect how execution time of a blind program may be affected by the building blocks of BEF, Equation \ref{eq:bitwise-total} shows the extrapolated polynomial of the total runtime consumed by the three building blocks of BEF with variables $a$, $b$, and $c$ denoting the number of times each algorithm is used. 

\begin{table}
  \centering
  \caption{Runtime Complexity for Blind Algorithms}
  \label{tab:blind-runtime}
  \begin{tabularx}{\columnwidth}{|X|X|}
    \hline
    \textbf{Blind Algorithm} & \textbf{Runtime ($m$ = number of bits)}\\
    \hline
    Blind Comparison & $O(m)$ \\
    \hline
    Blind Selection & $O(m)$ \\
    \hline
    Blind Ordering & $O(m)$ \\
    \hline
    Blind Min/Max & $O(n * m)$ \\
    \hline
    Blind Sorting & $O(n^2 * m)$ \\
    \hline
  \end{tabularx}
\end{table}

\begin{align}
    f(x) &= 52.2x - 30.29 \nonumber    \\
    g(x) &= 156.6x - 82.34 \nonumber   \\ 
    h(x) &= 12.79x + 492.3
    \label{eq:bitwise-blindalgo}
\end{align}

\begin{align}
    t(x, a, b, c) = a * f(x) + b * g(x) + c * h(x)
    \label{eq:bitwise-total}
\end{align}

Table \ref{tab:blindexecutiontime} shows the execution time and memory usage of the core algorithms of BEF using 4-bit binary data as inputs (refer to Table \ref{tab:blind-runtime} for runtime complexity of these algorithms). As it can be seen, blind selection and blind ordering take considerably longer to execute than blind comparison. However, these algorithms with blind selection in particular are integral to blind programming, with branching, decision-making, and control structure executions all relying on blind selection. The difference in memory usage between these three algorithms is relatively low with all of them staying within a hundredth of a megabyte within each other. In addition, it was observed that key generation, setup, and the encryption process of FHE consumed $88.334$ megabytes of memory, leading to our conclusion that while the bit-wise blind operations themselves don't add a significant amount of memory usage, the overwhelming memory requirements of FHE alone overshadow much of the rest of memory complexity for individual algorithms.

\begin{table}[htbp]
  \centering
  \caption{Blind Algorithms Execution Time and Memory Usage with 4-bit Inputs}
  \label{tab:blindexecutiontime}
  \begin{tabularx}{\columnwidth}{|X|X|X|X|}
    \hline
    \textbf{Algorithms} & \textbf{Execution Time (seconds)} & \textbf{Memory Usage (Mb)} \\
    \hline
    \textbf{Blind Comparison} & 163.3 & 88.448 \\
    \hline
    \textbf{Blind Selection} & 497.8 & 88.456 \\
    \hline
    \textbf{Blind Ordering} & 543.2 & 88.484 \\
    \hline
  \end{tabularx}
\end{table}

\subsection{Non-interactive Decision Trees with BEF}
\begin{figure}[htbp]
  \centering
  \includegraphics[width=\columnwidth]{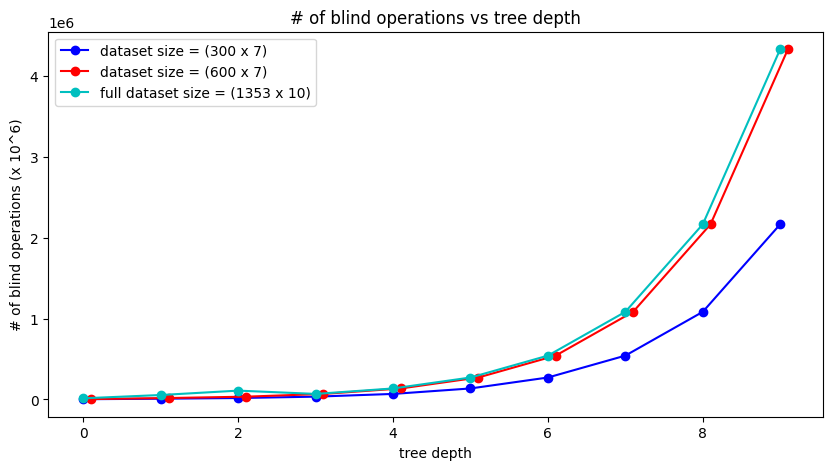}
  \caption{Number of total blind operations vs tree-depth with dataset size (n x m)}
  \label{fig:num-blind-op}
\end{figure}

\begin{figure}[htbp]
  \centering
  \includegraphics[width=\columnwidth]{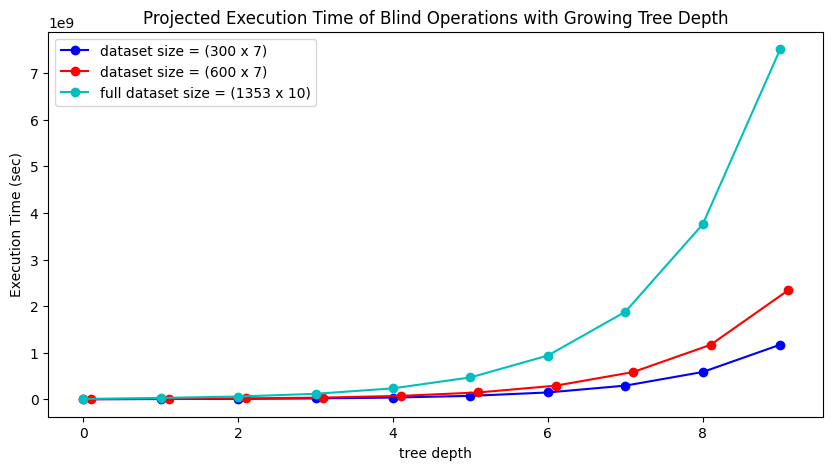}
  \caption{Execution time of blind operations vs tree-depth with dataset size (n x m)}
  \label{fig:time-blind-op}
\end{figure}

With BEF being the core building-block of non-interactive, encrypted training and prediction of decision trees, analysis was done to examine the footprint of blind operations in the context of decision trees. Table \ref{tab:number-blind-op} shows the number of times each blind operation is used at different points in decision tree training. Building from Table \ref{tab:number-blind-op}, Figure \ref{fig:num-blind-op} shows the total number of blind operations required in decision tree training with growing tree-depth and varying dataset size $(n * m)$ where $n$ is the number of rows (entries) and $m$ is the number of attributes (features). Using Equations \ref{eq:bitwise-blindalgo} and \ref{eq:bitwise-total}, the projected execution time of all blind operations in decision tree construction on various dataset sizes is shown in Figure \ref{fig:time-blind-op}.

\begin{table}[htbp]
  \centering
  \caption{Number of Blind Algorithms used (n = \# rows, m = \# attributes in dataset)}
  \label{tab:number-blind-op}
  \begin{tabularx}{\columnwidth}{|X|X|X|X|X|}
    \hline
    \textbf{Decision Tree Operations} & \textbf{Blind Comparison} & \textbf{Blind Selection} & \textbf{Blind Ordering} \\
    \hline
    \textbf{Feature Selection} & $m$ & $n * m$ & $3 * m$ \\
    \hline
    \textbf{Partition and Growing (per child node)} & $m$ & $n * 2m$ & $3 * m$ \\
    \hline
  \end{tabularx}
\end{table}

The effectiveness of BEF-enabled decision tree modeling was evaluated using the Phishing Website Detector dataset. This dataset contains 1353 data entries with 10 binary attributes. The attributes of this dataset include the existence of pop-up windows, validity of website IP address, and Server Form Handler (SFH). Using these features, a binary target (label) vector is used to determine whether or not a particular website with given attributes is a phishing scheme or a legitimate website. In order to demonstrate training in a timely manner, 300 stratified data entries were used with 7 attributes deemed most relevant. The 7 attributes were chosen using scikit-learn's SelectKBest algorithm using F-value classification as the scoring function. Since our work is the first of its kind handling non-interactive training of a machine learning model with FHE, it cannot be directly compared to other works that employ IRDE since they offload computational tasks to the client, thus making the server appear more efficient.
\\
Rust library for TFHE \cite{TFHE-rs} is used for the code implementation of the training and prediction paradigm presented in this paper \cite{myCode}. As of March 2023, \cite{TFHE-rs} began supporting encrypted comparison and min/max operations using similar underlying techniques as Algorithm \ref{algo:blindcomparison} and Algorithm \ref{algo:swapping} and these functions from \cite{TFHE-rs} are used for implementation of this work. While this shows that BEF algorithms are not novel on their own, their use in enabling non-interactive PPML, particularly in decision tree training, remain to be the first of its kind to the best of our knowledge.
\\
As this demonstration simply aims to showcase a methodology for non-interactive decision tree training, a primitive feature selection algorithm that measures class loss was implemented in place of gini impurity at the cost of performance in order to alleviate computational complexity. While BEF is not incompatible with more sophisticated feature scoring algorithms for feature selection, it was determined that the added computational complexity would be unnecessary for this demonstration as non-interactive calculation of these feature scoring functions have been shown in the past \cite{akavia2022privacy}. Rather, in feature selection, our contribution is showcasing the blind selection of the correct feature based on these calculated feature values in a non-interactive manner, which previously required IRDE \cite{akavia2022privacy}. 
\\
In addition to the encrypted model, a parallel implementation was developed in Python, employing the same BEF-enabled decision tree modeling approach, but without incorporating FHE. This plaintext (non-encrypted) implementation aims to benchmark the fundamental accuracy and viability of BEF, notwithstanding the inherent constraints of FHE. To further validate this approach, a comparative study using Scikit-learn's decision tree model was conducted on the same dataset and tree-depths. 
\\
The preliminary results are shown in Table \ref{tab:accuracyChart}. Validation accuracy for both plaintext and encrypted models were averaged after performing prediction on 30 batches of data. In addition, t-stat of $2.97\times 10^{-15}$ derived from p-value test using the prediction accuracy at tree-depth of 3 indicate that the plaintext and encrypted models are statistically consistent.
\\
These results underscore the viability and potential of BEF-enabled decision tree modeling within the scope of FHE's limitations. While accuracy is not the only metric critical in evaluating model performance, it does satisfy and validate the aim of this paper in showcasing that PPML models using BEF can be made trainable without IRDE within the scope of FHE's current computational limitations. While BEF enabled decision tree training is far from the state-of-the-art in runtime or memory usage, its novel capacity to perform training entirely on the server without IRDE opens the door to massive computational resources such as distributed computing grids or volunteer computing that cannot accommodate real-time IRDE between the client and computing entities. A more comprehensive analysis on time and memory usage for training and prediction phases is detailed in the Appendix.
\\
With this work showcasing that BEF can be deployed to successfully train privacy-preserving machine learning models without IRDE, future work in this direction will focus on optimization of performance metrics as well as efficiency in server-side computing as this paradigm is highly parallelizable.

\begin{table}[htbp]
  \centering
  \caption{Validation accuracy of plaintext and encrypted models at 95\% Confidence Interval with Scikit-Learn Decision Tree Classifier as Benchmark}
  \label{tab:accuracyChart}
  \begin{tabularx}{\columnwidth}{|X|X|X|X|}
    \hline
    \textbf{Tree Depth (d)} & \textbf{Plaintext BEF Accuracy} & \textbf{SK-Learn Decision Tree Accuracy} & \textbf{Encrypted BEF Accuracy} \\
    \hline
    d = 1 & $47.9\% \pm 5.1\%$ & 54.3\% & $42.9\% \pm 4.8\%$ \\
    \hline
    d = 2 & $62.3\% \pm 7.0\%$ & 62.3\% & $51.3\% \pm 6.2\%$ \\
    \hline
    d = 3 & $63.3\% \pm 5.2\%$ & 64.6\% & $63.3\% \pm 5.6\%$ \\
    \hline
  \end{tabularx}
\end{table}

\section{CONCLUSION}
This work proposed a novel framework for performing privacy-preserving training and prediction with FHE. The proposed solution leverages the Blind Evaluation Framework (BEF) with no Interactive Rounds of Decryption and Evaluation (IRDE) or communication during both training and prediction. In addition, the framework keeps the architecture of the system streamlined by keeping with a client-server model without additional computing parties involved. While BEF was used to demonstrate encrypted decision tree modeling in this work, the building blocks of BEF on their own are made generic and abstract to allow for flexible deployment in other applications that can benefit from blind programming. BEF presents a first step in achieving true outsourced computation of PPML with FHE.
\\
The preliminary results gathered in this work confirm that this framework presents a feasible approach to privacy-preserving decision trees, and open new possibilities for privacy-preserving machine learning as a whole. While approaches using IRDE can claim better performance metrics and computational efficiencies in their PPML implementations, such claims are misleading as portions of their programs must be executed in decrypted, plaintext form. While BEF in its current form has many limitations, it presents a programmatic framework that can enable truly encrypted PPML. 
\\
The aim for future work is two-fold. First, this research effort aims to address computational complexities and costs of this framework with parallelization techniques and more powerful computing units to make this paradigm trainable at a larger scale. Second objective is to introduce optimization efforts for performance metrics such as replacing primitive feature selection techniques with state-of-the-art algorithms and introducing cross-validation techniques into the paradigm.

\clearpage
\section*{APPENDIX}
\begin{figure}[htbp]
  \centering
  \includegraphics[width=\columnwidth]{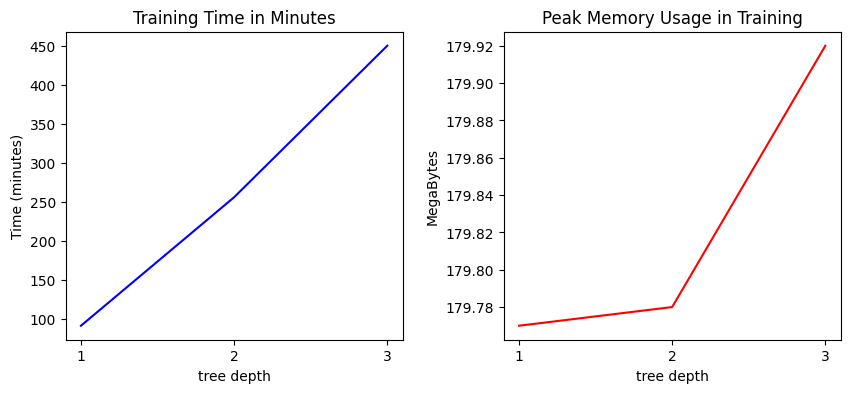}
  \caption{Training Time (left) and Memory Usage (right)}
  \label{fig:training-time}
\end{figure}

\begin{figure}[htbp]
  \centering
  \includegraphics[width=\columnwidth]{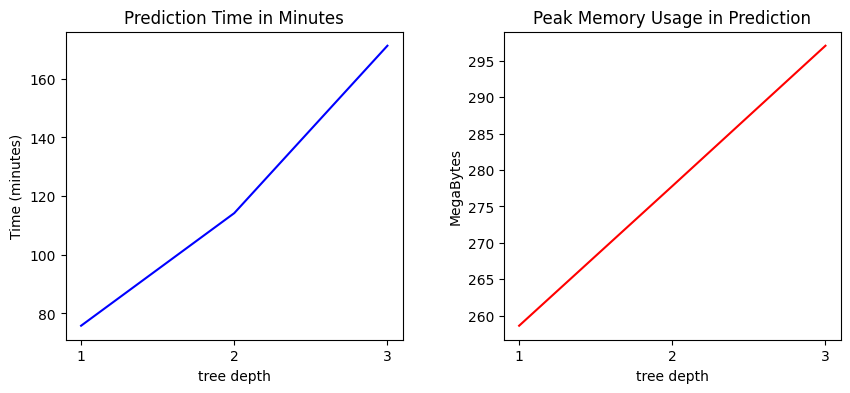}
  \caption{Prediction Time (left) and Memory Usage (right)}
  \label{fig:pred-time}
\end{figure}

\begin{figure}[htbp]
  \centering
  \includegraphics[width=\columnwidth]{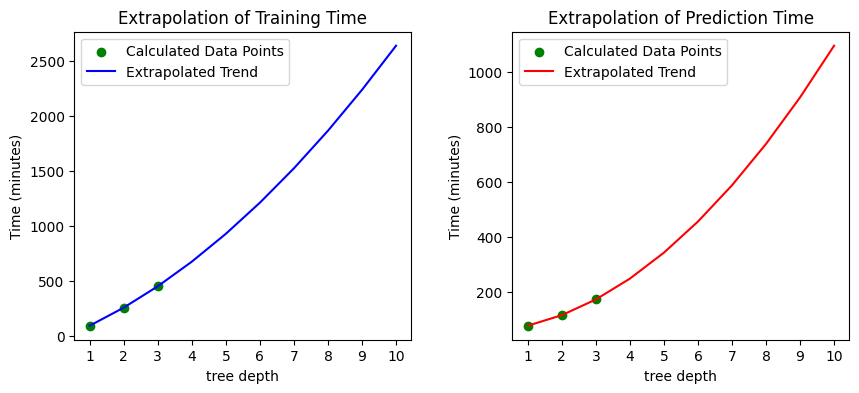}
  \caption{Extrapolation of Training (left) and Prediction Time (right)}
  \label{fig:extrapolation}
\end{figure}

Figure \ref{fig:training-time} shows the training time (in blue) and memory usage (in red) of the encrypted decision tree training with BEF with the x-axis indicating tree-depth. As shown in Table \ref{tab:blindexecutiontime}, the overwhelming memory requirements for FHE overshadow the additional memory overhead of different blind algorithm executions for training. Even though this figure shows exponential growth in memory usage with increasing tree-depths, the overall difference remains to be less than 0.5 Megabytes between tree-depth of 1 and tree-depth of 3. Time and memory usage for the prediction phase is shown in Figure \ref{fig:pred-time}.
\\
Extrapolating from the timing data from training and prediction, Figure \ref{fig:extrapolation} shows the projection of growth in training and prediction time with increasing tree-depth. The green dots indicate the calculated training and prediction times and the blue and red lines show the extrapolation based on the trajectory of these points for training and prediction, respectively. 

\bibliographystyle{IEEEtran}
\bibliography{PPDT}

\vfill\pagebreak

\end{document}